\documentclass[prb,twocolumn,aps,showpacs,fixfloats]{revtex4}
\usepackage{graphicx}
\usepackage{bm}
\usepackage{amsmath,amssymb}
\usepackage{subfigure}
\usepackage{float}
\usepackage{latexsym}
\usepackage{epstopdf}
\usepackage{color}
\usepackage{enumerate}
\usepackage{pdfpages}

\newcommand{\R}{\mathcal{R}}
\newcommand{\Rs}{\widetilde{\mathcal{R}}}

\begin{document}

\title {Manifestation of two-channel nonlocal spin transport in the shapes of the Hanle curves }

\author{R. C. Roundy$^{1}$, M. C. Prestgard$^{2}$, A. Tiwari$^{2}$, E. G. Mishchenko$^{1}$, and M. E. Raikh$^{1}$ }

 \affiliation{$^{1}$Department of Physics and
Astronomy, University of Utah, Salt Lake City, UT 84112, USA \\
$^{2}$Department of Materials Science and Engineering, University of Utah, Salt Lake City, Utah 84112, USA
}



\begin{abstract}
 Dynamics of charge-density  fluctuations in a system of two
tunnel-coupled wires  contains two diffusion modes
with dispersion $i\omega=Dq^2$ and $i\omega =Dq^2+\frac{2}{\tau_t}$, where $D$ is the diffusion
coefficient and $\tau_t$ is the tunneling time between the wires.
The dispersion
of corresponding spin-density modes depends on magnetic field as a
result of spin precession with Larmour frequency, $\omega_L$.
The presence of two modes affects the shape of the Hanle curve describing
 the spin-dependent resistance, $R$, between ferromagnetic strips covering the non-magnetic wires. We demonstrate that the relative shapes of the $R(\omega_L)$-curves, one measured within the same wire and the other measured between the wires,
 depends on the ratio $\tau_t/\tau_s$, where $\tau_s$ is
 the spin-diffusion time. If the coupling between the wires is local, i.e. only at the point $x=0$, then the difference of the shapes of intra-wire and inter-wire Hanle curves reflects
 the difference in statistics of diffusive trajectories which ``switch" or do not switch near $x=0$. When one of the coupled wires is bent into a loop with a radius, $a$,
 the shape of the Hanle curve reflects the
 statistics of random walks on the loop. This statistics is governed by the dimensionless parameter,
 $\frac{a}{\sqrt{D\tau_s}}$.

\end{abstract}

\maketitle

\begin{figure}
\includegraphics[width=90mm]{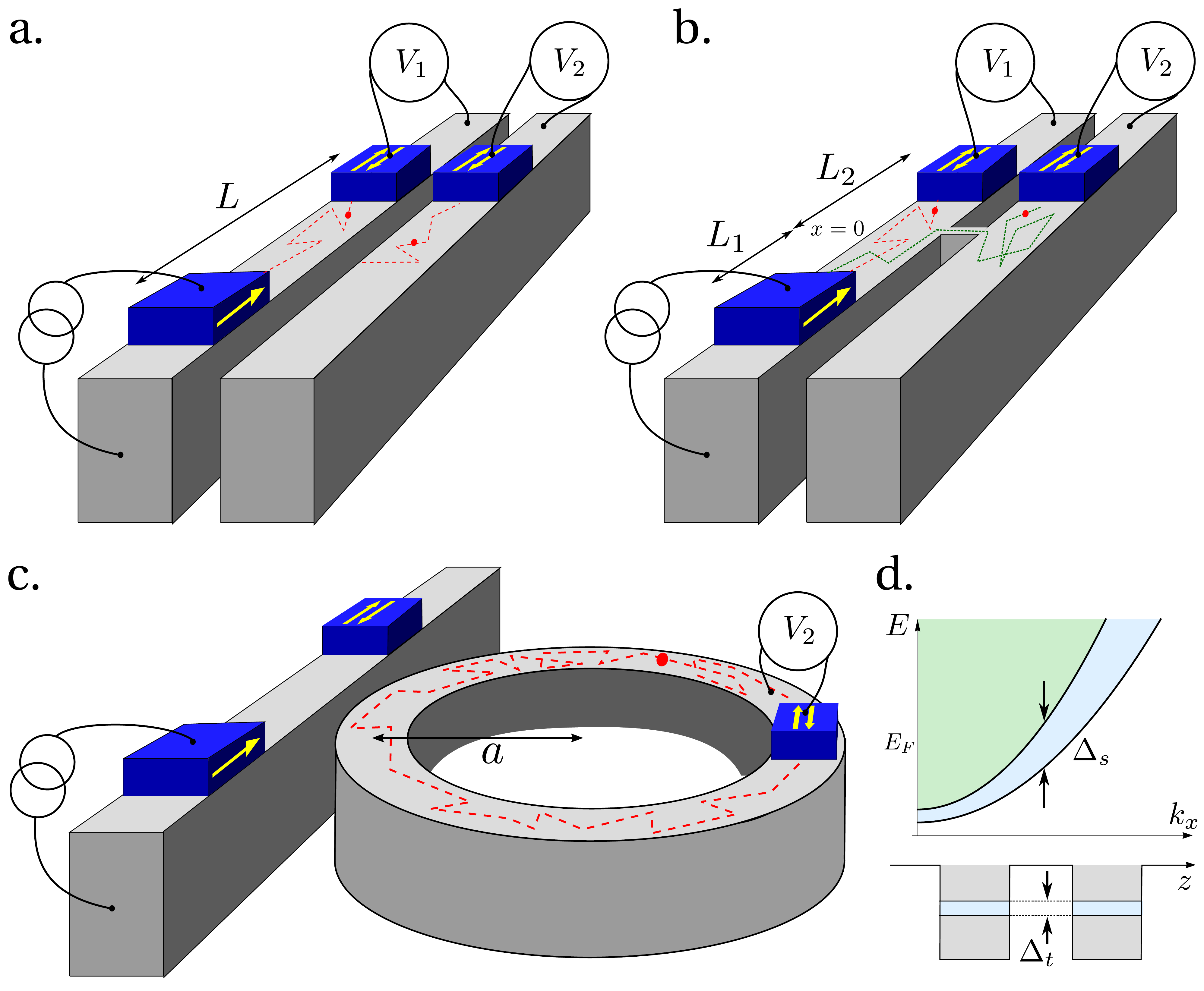}
\caption{(Color online) (a) Two-channel spin-transport device. The injector is located in the left channel. Two detectors in the left and right channels are located at the same distance, $L$, from the injector. An electron reaches the first detector by diffusion and the second detector by a combined
diffusion-tunneling process; (b) In contrast to (a) two channels are coupled {\em locally} at the
point $x=0$; (c) the second wire is bent into a loop. Electron diffusion trajectories encircle the loop several times before the spin polarization is ``forgotten"; (d) two branches, $E(k_x)$, of the energy spectrum of the tunnel-coupled wires. At small momenta, the splitting,  $\Delta_t$, is determined by tunneling, while at large $k_x$ the splitting $\Delta_s$ is dominated by the spin-orbit coupling in the wires.}
\label{devices}
\end{figure}

\section{Introduction}

Spin-orbit interaction is the origin of spin dephasing
in semiconductors and metals. On a microscopic level, a finite
spin-relaxation time, $\tau_s$,  results from
the momentum-dependent spin-orbit term in the Hamiltonian of a free electron in
combination with scattering-induced momentum relaxation.\cite{dyakonov-perel} In metals, the origin of spin dephasing is spin-dependent impurity scattering.

First experimental studies\cite{Parsons1969,Hanle1970,Hanle1971}  of spin relaxation in semiconductors were carried out  more than four decades ago. They were based on the notion that if the photoexcited electron has its spin
pointing along the $x$ --axis, then in magnetic field, $\omega_L$,  directed along the $z$ --axis, the projection, $S_x(t)$, evolves as  $S_x(t)=\cos \omega_Lt \exp{(-t/\tau_s)}$ which is the result of the Larmour precession. Since the time between generation and recombination is much longer than $\tau_s$, so that the spin evolution is completed by the moment of recombination, then the polarization of the luminescence is proportional to $\int_0^\infty dt S_x(t)$, i.e.
\begin{equation}
\label{optical}
{\cal P}(\omega_L)=\frac{{\cal P}(0)}{1+\omega_L^2\tau_s^2}.
\end{equation}
Numerous experimental measurements reported to date can be fit very accurately with Lorentzian Hanle profile Eq. (\ref{optical}), and when they do not, see e.g. Ref. \onlinecite{trion}, the deviations reflect the peculiarity of the recombination process.

In the pioneering papers Refs. \onlinecite{Silsbee1985}, \onlinecite{Silsbee1988} it was demonstrated
that, aside from optics,
the underlying physics of spin relaxation manifests itself in transport
experiments. The structure fabricated and measured in Ref. \onlinecite{Silsbee1985} represented an aluminum wire with two cobalt ferromagnetic strips on the top.
The first strip, injector, played the role of circular-polarized
excitation light in optics, in the sense, that it supplied spin-polarized electrons
into the wire.  Correspondingly, the second strip, the detector, imitated the
analyzer of the emitted light. The characteristic
measured in Ref. \onlinecite{Silsbee1985} was the nonlocal resistance, $R$, which is the ratio of the voltage, generated between the channel and  detector, to the current passed through the injector into the channel.

Similar to polarization of luminescence, ${\cal P}(\omega_L)$,
the nonlocal resistance is suppressed with external field, $\omega_L$. There is, however, a fundamental difference between
the dependencies ${\cal P}(\omega_L)$ and $R(\omega_L)$.
This difference stems from the fact that, in addition to the Larmour precession, formation of nonlocal resistance involves
diffusion of carriers over the distance, $L$, between injector and detector. This diffusion is routinely incorporated into the theory by multiplying $\cos\omega_Lt\exp(-t/\tau_s)$ by a diffusion propagator, $P_L(t)$, and only subsequently integrating over time.
In one dimension,  $P_L(t)$  has the form
\begin{equation}
\label{propagator}
P_L(t)=\frac{1}{\left(4\pi Dt\right)^{1/2}}\exp\left(-\frac{L^2}{4Dt}\right),
\end{equation}
where $D$ is the diffusion coefficient.

The nonlocal resistance, $R(\omega_L)$, calculated
with the help of propagator Eq. (\ref{propagator}),
is also called the Hanle curve in the literature.
The expression for $R(\omega_L)$ contains two
unknown parameters, $\tau_s$ and $D$.
Still it appears that
the scores of
experimental data accumulated to date can be fitted very accurately with this expression. It is apparent that the shapes of $R(\omega_L)$ is different for ``short"  $L\ll (D\tau_s)^{1/2}$ and long   $L\gg (D\tau_s)^{1/2}$ samples.
This difference in shapes was pointed out already in the seminal paper Ref. \onlinecite{Silsbee1985}, where the two samples measured had the lengths $L=50$~$\mu$m and $L=300$~$\mu$m.

Experimental studies of nonlocal spin transport became a hot topic  in 2001 when the measurements of $R(\omega_L)$ were reported\cite{vanWeesPioneering} for  small samples with $L\sim 0.5$~$\mu$m. Small structures are appealing for information-storage applications. Indeed,
the effect of sign reversal of nonlocal resistance upon reversal of the magnetization of the detector allows one to view the
detector as an element of information storage.
For this reason, the $R(\omega_L)$ --measurements in spin transport were carried out
since 2001 on various structures with
$L$ in the micrometer range and with materials of  non-magnetic channels ranging from Si and GaAs, see e.g.
Refs. \onlinecite{AppelbaumSi} and \onlinecite{Crowell2007},
to graphene\cite{graphene2007} and organic materials.\cite{HanleOrganicJapanese}

It turns out that the scope of experimental results on nonlocal resistance are described by the drift-diffusion theory with remarkable precision and including finest details, see e.g. Refs.
\onlinecite{Ge2014, Wunderlich, graphene2013, graphene2014}.
 For example,
in long samples $L\gg \left(D\tau_s\right)^{1/2}$,
the theory predicts several zeros in $R(\omega_L)$ dependence.
The origin of these zeros is that during the time $L^2/D$ of travel between the injector and detector the spin can precess
by $2\pi$, $4\pi$, and so on. Clearly, the values of
$R(\omega_L)$ between these zeros fall off dramatically.
Then the number of the zeros observed in experimental $R(\omega_L)$ attests to the accuracy with which the theory captures the process of spin transport.
Usually, only the first zero is resolved in experiment. However, the very recent data in Ref. \onlinecite{graphene2014} exhibits the second zero as well.

\begin{figure}
\includegraphics[width=77mm]{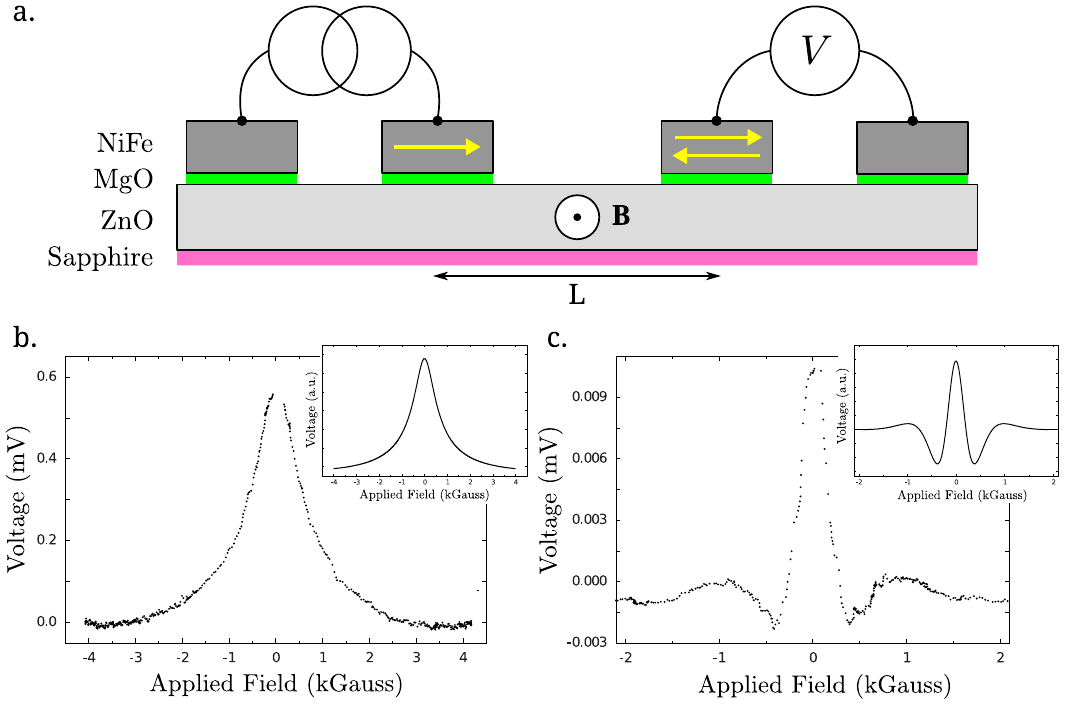}
\caption{[Color Online] (a) Schematic view of a four-terminal device used for nonlocal spin-transport measurements\cite{Megan}. Epitaxial ZnO film  of a thickness $200$~nm, deposited on a sapphire substrate, is spaced from a NiFe layer by a thin MgO barrier; (b) and (c) are the Hanle curves measured for the channel length $L=90$~nm and $L=650$~nm, respectively. The insets show the theoretical
fits plotted from Eqs. (\ref{R}), (\ref{integralF}), (\ref{f(y)}), and (\ref{complex}).
The values $\tau_s$ used for both fits are the same, so that the dimensionless lengths ${\cal L}=L/\sqrt{4D\tau_s}$ differ $7$ times.        }
\label{tiwari-data}
\end{figure}

To illustrate the accuracy with which the drift-diffusion theory works for
spin-transport devices with variable channel length, in Fig.~\ref{tiwari-data}(b),~(c) we
plot the Hanle shapes measured for two devices fabricated from  epitaxial ZnO films.
Both devices were fabricated under the same conditions\cite{Megan}, which included  pulsed-laser deposition
of ZnO onto a sapphire substrate,
 deposition of a thin barrier layer of MgO on top, and, finally,  the deposition of  NiFe electrodes  using photolithography and e-beam evaporation. The only difference between the  two devices 
 was the distance between the contacts ($L=90$~nm and $L=650$~nm) in the four-probe structure, see Fig. \ref{tiwari-data}(a). We see that the seven-fold increase in $L$ changes the shape dramatically, in {\em quantitative} agreement with predictions of the drift-diffusion theory.

The fact that the measured $R(\omega_L)$
is so accurately described by the drift-diffusion
theory suggests that the shape of the Hanle curve is the
characteristics of the spin transport in non-magnetic channel only, and
is not affected by the details of injection and detection.
It also suggests that description of electron diffusion
paths as purely one-dimensional is surprisingly adequate.
In fact, in experimental geometries, the length, $L$, of the
channel does not exceed significantly the channel width.

This motivated us to study theoretically the shapes of the
Hanle curves in geometries when the transport between the injector
and detector does not reduce to a 1D random walk. The results of this study are reported
in Sections II and III, where we consider the
spin transport along two parallel tunnel-coupled wires and the
transport in the special case when one of the wires is bent into
a loop. Our objective was to find out  whether the
statistics of diffusion paths, specific for quasi 1D geometries,
can be distinguished in nonlocal spin-transport measurements.
Our main message is that, the difference of the Hanle curves
measured with two detectors, one located in the same wire as injector
and the other located in the neighboring wire,
reflects the peculiar statistics of the diffusion
paths in a coupled system.

%
%
%
%
%
%

\section{Spin-density fluctuations in tunnel-coupled wires}

%

Fluctuations of electron densities, $n_1(x,t)$ and $n_2(x,t)$, in a system of two coupled wires satisfy the system of equations
\begin{align}
\label{charge}
\frac{\partial n_1}{\partial t} &= D \frac{\partial^2 n_1}{\partial x^2} - \frac{1}{\tau_t} (n_1 - n_2), \nonumber\\
\frac{\partial n_2}{\partial t} &= D \frac{\partial^2 n_2}{\partial x^2} - \frac{1}{\tau_t} (n_2 - n_1),
\end{align}
where $\tau_t$ is the inter-wire tunneling time. We assume that the wires are
disordered so that $\tau_t \gg \tau$, where $\tau$ is the disorder--scattering time.
The latter condition implies that the tunnel splitting, $\Delta_t$, of the
spectra of the wires in the absence of disorder is much smaller than $\tau_t^{-1}$.
In this limit, the expression for $\tau_t$ reads\cite{Rosch}
\begin{equation}
\tau_t=\frac{1}{\Delta_t^2\tau}.
\end{equation}
Equations Eq. (\ref{charge}) give rise to two diffusion modes with dispersions
\begin{equation}
i\omega =Dq^2, ~~~~i\omega=Dq^2+\frac{2}{\tau_t}
\end{equation}
corresponding
to symmetric and antisymmetric distributions of densities,
respectively.
So that the actual distributions $n_1(x,t)$ and $n_2(x,t)$ are
linear combinations of the two modes.
If electrons are injected into the first wire at $x=0$,  the combinations satisfying the initial conditions
\begin{equation}
n_1(x,0)=\delta(x),~~~~  n_2(x,0)=0.
\end{equation}
are the sum and the difference of two diffusion modes
\begin{align}
\label{n1,n2}
n_1(x, t) &= \frac{1}{2}~ P_{x}(t)\left[ 1 + e^{-2t/\tau_t}\right],\nonumber \\
n_2(x, t) &= \frac{1}{2}~ P_{x}(t)\left[ 1 - e^{-2t/\tau_t}\right],
\end{align}
where the diffusion propagator $P_{x}(t)$ is defined by Eq.~(\ref{propagator}).

To describe the nonlocal resistance  we need the expressions for
the spin densities, ${\bm S}_1(x,t)$ and  ${\bm S}_2(x,t)$, similar to Eq.~(\ref{n1,n2}). The system
of coupled equations for ${\bm S}_1(x,t)$,  ${\bm S}_2(x,t)$ has the form
\begin{align}
\label{general}
\frac{\partial {\bm S}_1}{\partial t} &= {\bm \omega}_L \times {\bm S}_1 - \frac{{\bm S}_1}{\tau_s}
+ D \frac{\partial^2 {\bm S}_1}{\partial x^2} - \frac{1}{\tau_t} \left( {\bm S}_1 - {\bm S}_2 \right), \nonumber \\
\frac{\partial {\bm S}_2}{\partial t} &=  {\bm \omega}_L \times {\bm S}_2 - \frac{{\bm S}_2}{\tau_s}
+ D \frac{\partial^2 {\bm S}_2}{\partial x^2} - \frac{1}{\tau_t} \left( {\bm S}_2 - {\bm S}_1 \right),
\end{align}
and differs from the corresponding equations
Eq. (\ref{charge}) describing the charge-density fluctuations in two aspects: both ${\bm S}_1$ and  ${\bm S}_2$, precess in magnetic field, ${\bm\omega}_L$, and both decay during the spin relaxation time $\tau_s$
\begin{equation}
\tau_s=\frac{1}{\Delta_s^2\tau},
\end{equation}
where $\Delta_s$ is the spin-orbit splitting of the spectrum in each wire in the absence of disorder.
Note that, while both $\tau_t$ and $\tau_s$ contain scattering time, the ratio $\tau_s/\tau_t$ does not contain disorder, i.e. it is a characteristics of clean wires.
The term coupling the wires in the system Eq. (\ref{general}) has the same form as in the system
Eq. (\ref{charge}) since tunneling conserves the spin.

Without boundary conditions, the system Eq. (\ref{general}) defines four modes
\begin{equation}
i \omega = - Dq^2 - \frac{1}{\tau_s} \pm~i \omega_L,~~i \omega= - Dq^2 - \frac{1}{\tau_s}-
\frac{2}{\tau_t} \pm~ i \omega_L,
\end{equation}
of which the first two correspond to symmetric and the second two to the antisymmetric spin-density fluctuations. With boundary conditions, the solution of the system  Eq. (\ref{general}) can be expressed in terms of the solution Eq. (\ref{n1,n2}) of the system Eq. (\ref{charge}) as follows

\begin{equation}
\label{bigS}
{\bm S}_1(x,t)={\bm s}(t) n_1(x,t),~~~{\bm S}_2(x,t)={\bm s}(t)n_2(x,t),
\end{equation}
where the function ${\bm s}(t)$ satisfies the conventional equation of spin dynamics
\begin{equation}
\label{dynamics}
\frac{d {\bm s}}{dt}={\bm \omega}_L\times {\bm s}-
\frac{{\bm s}}{\tau_s}.
\end{equation}

\section{Nonlocal resistances}

The initial condition to Eq. (\ref{dynamics}) is set by the direction of polarization of the injector.
We assume that ${\bm s}(0)$ is directed along the $x$ --axis.

As it is illustrated in Fig. \ref{devices}, there are two nonlocal resistances: $R_{11}(\omega_L)$
is the resistance measured by the detector within the same wire, $1$, where polarized electrons
are injected, and $R_{12}$ is the resistance measured by the detector that covers the wire $2$.
Within a prefactor they are given by
\begin{equation}
\label{definition}
\R_{11}=R_0 L \int\limits_0^{\infty}\frac{dt}{\tau_s}~ S_{1x}(L,t),~~\R_{12}=R_0 L \int\limits_0^{\infty}\frac{dt}{\tau_s}~ S_{2x}(L,t).
\end{equation}
In Eq. (\ref{definition}) it is  implicit that the magnetization of the detector is also along the $x$-- axis. In some experiments, say Ref. \onlinecite{Crowell2005}     , the spin transport was studied for polarization of the detector along the $y$ --axis. The corresponding expression for nonlocal resistance reads
\begin{equation}
\label{definition1}
\Rs_{11}=R_0 L \int\limits_0^{\infty}\frac{dt}{\tau_s}~ S_{1y}(L,t),~~\Rs_{12}=R_0 L \int\limits_0^{\infty}\frac{dt}{\tau_s}~S_{2y}(L,t).
\end{equation}

Our goal is to find the expressions for $\R_{11}(\omega_L)$ and $\R_{12}(\omega_L)$ for two tunnel-coupled wires. One can see that the coupling strength, $\tau_t^{-1}$, enters into the formulas for $\R_{11}(\omega_L)$, $\R_{12}(\omega_L)$ through the last terms in Eq. (\ref{charge}). These terms
decay exponentially with time. We also notice that ${\bm s}(t)$ which satisfies Eq. (\ref{dynamics})
is also an exponential function of time. This observation allows one to express $\R_{11}(\omega_L)$, $\R_{12}(\omega_L)$ with tunneling through nonlocal resistance, $R(\omega_L)$,  {\em in the absence of tunneling}.

Setting $\tau_t=\infty$ and substituting
\begin{equation}
\label{littleS}
{\bm s}(t) = e^{-t/\tau_s}\Bigl({\bm i}\; \cos \omega_L t + {\bm j}~\sin \omega_L t \Bigr)
\end{equation}
into Eq. (\ref{bigS}) and subsequently into Eq. (\ref{definition})
 we restore a standard expression for the Hanle profile of a single channel
\begin{equation}
\label{R}
R(\omega_L, \tau_s) = R_0 F(\omega_L, \tau_s)
\end{equation}
where the dimensionless function $F(\omega_L, \tau_s)$ is defined as
\begin{equation}
\label{integralF}
F(\omega_L, \tau_s) = \frac{L}{\tau_s} \int\limits_0^\infty dt \cos \omega_L t~ e^{-t/\tau_s} P_L(t),
\end{equation}
so that $R_0$ has the dimensionality of the resistance.

Then, in terms of the function $R(\omega_L, \tau_s)$, the final result for nonlocal resistances can
be presented as
\begin{align}
\label{final}
\R_{11}=\frac{1}{2}\Bigl [R(\omega_L, \tau_s)+ \frac{{\tilde \tau}_s}{\tau_s} R(\omega_L,{\tilde\tau}_s)\Bigr],\nonumber\\
\R_{12}=\frac{1}{2}\Bigl [R(\omega_L, \tau_s)-\frac{{\tilde \tau}_s}{\tau_s} R(\omega_L,{\tilde\tau}_s)\Bigr],
\end{align}
where ${\tilde\tau}_s$ is an effective spin relaxation time
\begin{equation}
\label{tilde}
{\tilde\tau}_s =\frac{\tau_s\tau_t}{2\tau_s+\tau_t}
\end{equation}
which includes tunneling an is shorter than $\tau_s$. Modifications of the Hanle
curves due to tunneling are analyzed below.

\begin{figure}
\includegraphics[width=77mm]{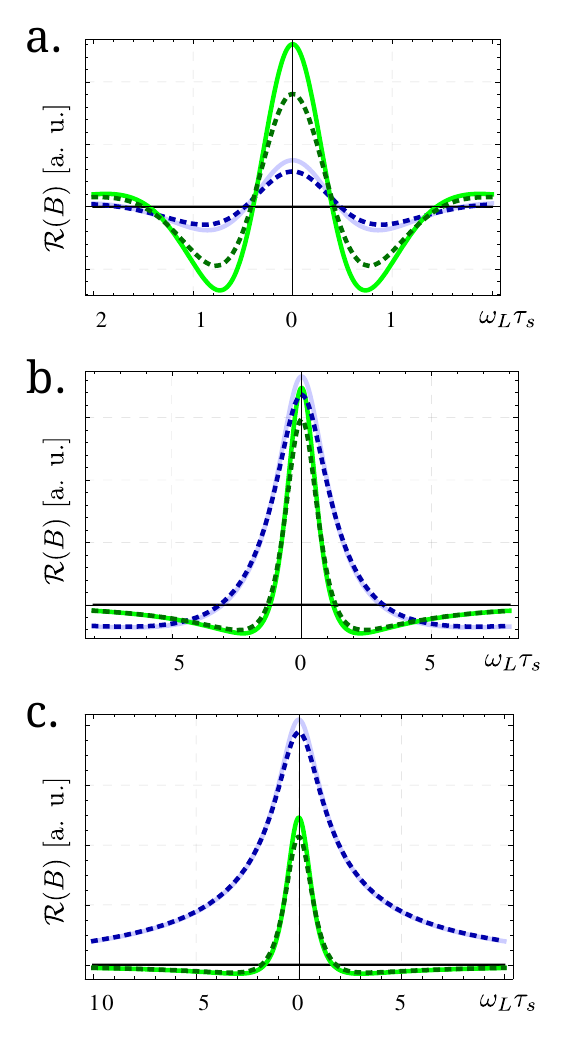}
\caption{[Color online]
Hanle curves ${\cal R}_{11}(\omega_L)$ (blue) and  ${\cal R}_{12}(\omega_L)$ (green) measured by two detectors, one located in the same wire as injector and the located in the neighboring wire.  Dashed curves are plotted directly from Eq. (\ref{final}) using the definition Eq. (\ref{R}); the solid curves are plotted from the assymptotic expansions, $\R_{11}$ is
given by Eq. (\ref{integralF}) and $\R_{12}$ is given by  Eq. (\ref{approx-R12}).
In all graphs the tunneling time, $\tau_t$, is $10\tau_s$. The three panels correspond
to different dimensionless lengths, ${\cal L}=L/\sqrt{4D\tau_s}$:  ${\cal L}=3$ (a),
 ${\cal L} = 0.45$ (b), and  ${\cal L} = 0.1$ (c). All curves for ${\cal R}_{12}$ are multiplied by $\frac{\tau_t}{\tau_s}=10$.
}
\label{figTunnel}
\end{figure}

\subsection{Limiting cases}
As it was mentioned in the Introduction, the shape of the Hanle curve for a single wire
is governed by the dimensionless length
\begin{equation}
{\cal L}=\frac{L}{\sqrt{4D\tau_s}}.
\end{equation}

{\em i}.~~ It is apparent that when both wires are long, ${\cal L} \gg 1$, the shapes of
the curves ${\cal R}_{11}$ and ${\cal R}_{12}$ do not differ significantly,
since a typical electron will have enough time to tunnel before it reaches one of two detectors.

{\em ii}.~~ It is also obvious on general grounds that when, the tunneling time is much shorter than the spin--relaxation time, $\tau_t \ll \tau_s$, the nonlocal resistance ${\cal R}_{11}$ exceeds ${\cal R}_{12}$ only slightly.
This is because the electron gets equally distributed between the wires before the spin precession takes place.  Formally, this follows from Eqs. (\ref{final}) and (\ref{tilde}). In the limit $\tau_t \ll \tau_s$ one has ${\tilde \tau}_s \approx \tau_t/2$. The relative difference,
$\left({\cal R}_{11}-{\cal R}_{12}\right)/{\cal R}_{11}$, is of the order of $\left(\tau_t/\tau_s\right)^{1/2}$.

{\em iii}.~~ The opposite limit of weak tunneling between the wires is most insightful.
In this limit, we have $\tau_t \gg \tau_s$, so that only a small portion
of electrons injected in the first wire reach the detector in the second wire.
This means that ${\cal R}_{12}$, is
much smaller than ${\cal R}_{11}$. Formally, two terms in
Eq. (\ref{final}) for ${\cal R}_{12}$ nearly cancel each other.
However, ${\cal R}_{12}(\omega_L)$  possesses a distinctive shape.
To find this shape we expand Eq. (\ref{final}) with respect to $\tau_s/\tau_t$ and  get
\begin{equation}
\label{approx-R12}
\R_{12} 
\approx R_0 \frac{\tau_s}{\tau_t} G(\omega_L, \tau_s),
\end{equation}
where the function $G(\omega_L,\tau_s)$ is defined as
\begin{equation}
\label{g(y)}
G(\omega_L, \tau_s) = \frac{L}{\tau_s^2} \int\limits_0^\infty dt \; t
\cos \omega_L t~ e^{-t/\tau_s} P_L(t).
\end{equation}
Analytical expressions for $F(\omega_L, \tau_s)$ and $G(\omega_L, \tau_s)$ for arbitrary length can be found
using the identities
\begin{align}
\label{f(y)}
\int\limits_0^{\infty}\frac{ds}{s^{1/2}}\exp{\Bigl[-\frac{1}{s}-ys\Bigr]} &=
\Bigl(\frac{\pi}{y}\Bigr)^{1/2}   \exp{\Bigl[-2y^{1/2}\Bigr]}, \\
\hspace{-7mm}\int\limits_0^{\infty}ds~ s^{1/2}\exp\left[-\frac{1}{s}-ys\right]
&=\frac{\pi^{1/2}}{2y^{3/2}}\left(1+2y^{1/2}\right)\exp\left[-2y^{1/2}\right],
\end{align}
and taking the absolute value and the phase of the complex argument, $y$, to be
\begin{equation}
\label{complex}
\vert y \vert ={\cal L}^2 \Bigl(1+\omega_L^2{\tau}_s^2\Bigr)^{1/2},~~~~\phi=\arctan \left(\omega_L{\tau}_s\right).
\end{equation}
In Fig. \ref{figTunnel} we plot these functions which represent the Hanle curves for diagonal
and nondiagonal resistances for three domains of ${\cal L}$. It is seen, Fig. \ref{figTunnel}(a),  that for large length the shapes of both curves are identical.
The smaller is the length the more pronounced is the difference between ${\cal R}_{11}$
and  ${\cal R}_{12}$ behaviors. The ${\cal R}_{12}(\omega_L)$ curve is significantly narrower
than ${\cal R}_{11}(\omega_L)$ for small length, as it is seen in Fig. \ref{figTunnel}(c).
This narrowing originates from the extra factor, $t$, in the integrand of
Eq. (\ref{g(y)}) compared to Eq. (\ref{integralF}) and
can be qualitatively interpreted as follows.
In order to reach the detector in the second wire, injected electron diffuses along the first wire, tunnels into the second wire, and diffuses there.
Narrower shape indicates that reaching the detector in the second wire takes more time than
reaching the detector in the first wire to which electron simply diffuses.

{\em iv}.~~ Note that the limit of short wires allows a comprehensive analytical study to which we now turn.

In the limit of small wire length, ${\cal L} \ll 1$,
analytical expressions  for ${\cal R}_{11}$ and ${\cal R}_{12}$
can be obtained for arbitrary relation between $\tau_t$ and $\tau_s$.
In this limit, corresponding to $|y|\ll 1$ in  Eq. (\ref{f(y)}), the
expression for nonlocal resistances, $R(\omega_L, \tau_s)$ and ${\tilde R}(\omega_L, \tau_s)$ of an isolated wire simplify to
\begin{align}
\label{smallL}
R(\omega_L, \tau_s) &= \frac{R_0 {\cal L}}{\sqrt{2}}  \frac{\sqrt{\sqrt{1+\omega_L^2{\tau}_s^2}+1}}
{\sqrt{1+\omega_L^2{\tau}_s^2}}, \\
\label{smallLtilde}
{\tilde R}(\omega_L, \tau_s) &= \frac{R_0 {\cal L}}{\sqrt{2}}  \frac{\sqrt{\sqrt{1+\omega_L^2{\tau}_s^2}-1}}
{\sqrt{1+\omega_L^2{\tau}_s^2}}.
\end{align}
Substituting Eq. (\ref{smallL}) into Eq. (\ref{final}), we get
\begin{multline}
\label{plusminus}
\begin{pmatrix}
    \R_{11}(\omega_L) \\
    \R_{12}(\omega_L)
\end{pmatrix}
=\frac{R_0 {\cal L}}{2\sqrt{2}}  \\
\times \left[  \frac{\sqrt{\sqrt{1+\omega_L^2{\tau}_s^2}+1}}
{\sqrt{1+\omega_L^2{\tau}_s^2}} \right.
\left. \pm
\sqrt{\frac{{\tilde \tau}_s}{{\tau}_s}} \; \frac{\sqrt{\sqrt{1+\omega_L^2\tilde{\tau}_s^2}+1}}
{\sqrt{1+\omega_L^2\tilde{\tau}_s^2}} \right].
\end{multline}
The second term in Eq. (\ref{plusminus}) is responsible for the difference between $\R_{11}$ and $\R_{12}$. It is apparent that this difference is maximal when $\tilde{\tau}_s\approx \tau_s$, i.e. when the tunneling time is long. In the latter case we can simplify Eq.
(\ref{plusminus}) further by expanding with respect to $\tau_s/\tau_t$
\begin{align}
\label{asymptote}
\R_{12} &= \frac{R_0 {\cal L}}{2\sqrt{2}}
\frac{\tau_s}{\tau_t}
 \frac{1}{(1 + \omega_L^2 \tau_s^2)^{3/2}} \nonumber\\
 &\times \frac{\sqrt{1 + \omega_L^2 \tau_s^2}+ 1-\omega_L^2\tau^2}
 {\sqrt{\sqrt{1 + \omega_L^2 \tau_s^2}+ 1}}.
\end{align}
Now $\R_{12}$ is a function of a single argument, $\omega_L\tau_s$. We see that, while ${\cal R}_{11}$ falls off at large $\omega_L\tau_s$ as $\left(\omega_L\tau_s\right)^{-1/2}$,
the decay of ${\cal R}_{12}$
is much faster, as $\left(\omega_L\tau_s\right)^{-3/2}$. Besides, the distinctive feature
of ${\cal R}_{12}$ is that it passes through zero at $\omega_L\tau_s=\sqrt{3}$.
Overall, see Fig. \ref{figTunnel}, despite the length is small, the behavior of non-diagonal resistance ${\cal R}_{12}$  resembles the shape
of the Hanle curve for a long wire, ${\cal L} \gg 1$.

\section{Local coupling}
\begin{figure}
\includegraphics[width=77mm]{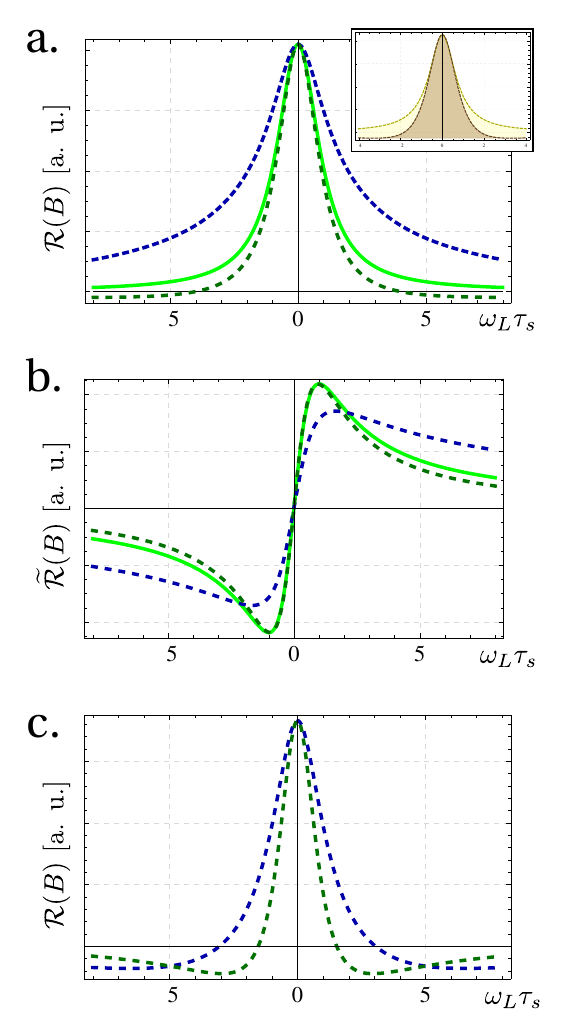}
\caption{[Color online] The difference in the diffusion trajectories in a single wire
and in two wires coupled via a bridge manifests itself in the shapes of the Hanle curves.
(a) The Hanle curves $\R_{11}(\omega_L)$ (blue) and $\R_{12}(\omega_L)$ (green, dashed)
are plotted from Eqs. (\ref{propto}) and (\ref{VeryLong}) for a small  dimensionless length ${\cal L} = L / (4 D \tau_s)^{1/2} = 0.1$.  The solid green curve is the Lorentzian
 asymptote, Eq. (\ref{surprising}). The inset shows a comparison of $\R_{12}(\omega_L)$
for the case of local tunneling (yellow), Eq.(\ref{surprising}), and for the case of homogeneous tunneling
(brown), Eq. (\ref{asymptote}); (b) same geometry as in (a). Non-diagonal components of nonlocal resistance corresponding to the perpendicular magnetizations of injector and detector are plotted;  (c) the Hanle curve $\R_{11}(\omega_L)$ (blue) and $\R_{12}(\omega_L)$ (green) are plotted for the dimensionless length  ${\cal L} = 0.45$.
The shapes are much closer than in (a).
}
\label{figH}
\end{figure}
A different arrangement of two coupled wires is shown in Fig. \ref{devices}(b). Electron injected into the first wire can cross into the second wire only through a narrow bridge at $x=0$. This means that, while ${\cal R}_{11}$ is constituted by all diffusive trajectories
in the first wire, the contribution to ${\cal R}_{12}$ comes from a subset of diffusive trajectories which visit the point of contact. More precisely, the bridge serves as
``weak" boundary condition for the diffusion equation.
We are going to study how this  modification of the diffusion due to
crossing into neighboring wire affects the shape of the Hanle curve, ${\cal R}_{12}$,
and compare the result with ${\cal R}_{12}(\omega_L)$ calculated for homogeneous tunneling
in the previous Section.

We assume that the coupling via the bridge is weak, so that the concentration $n_1(x,t)$
is given by $P_{x+L_1}(t)$. Presence of the bridge in the diffusion equation Eq. (\ref{charge}) for $n_2(x,t)$ is reflected as a source
\begin{equation}
\label{differential}
\frac{\partial n_2}{\partial t} - D \frac{\partial^2 n_2}{\partial x^2} =
\frac{l}{\tau_t} \delta(x) n_1(0, t)
\end{equation}
where $l\ll L$ represents the width of the bridge. The solution of Eq. (\ref{differential}) can be obtained in a standard way, e.g., by the Fourier expansion of both sides. The expression for $n_2(x,t)$ reads
\begin{equation}
n_2(x, t)  = \frac{l}{\tau_t} \int_0^t dt_1 P_{x}(t-t_1) n_1(0, t_1).
\end{equation}
Substituting the expression for $n_1(0,t)$ into the integrand, we cast  the final result in the form
\begin{equation}
\label{convolution}
n_2(x, t)=\frac{l}{\tau_t}\int_0^t dt_1\int_0^t dt_2 P_x(t_1)P_{L_1}(t_2)\delta(t_1+t_2-t),
\end{equation}
which is simply the convolution of two diffusion propagators.
Physically, the result Eq. (\ref{convolution}) is transparent. It expresses
the fact that, to get to the point $x$ in the second wire, electron first diffuses
from injector to the bridge and then from the bridge to the point $x$.

The form Eq. (\ref{convolution}) is convenient for the calculation of the nonlocal resistance ${\cal R}_{12}(\omega_L)$. Indeed, for this calculation one has to multiply $n_2(x,t)$ by
${\bm s}(t)$, given by Eq. (\ref{littleS}), and integrate over $t$, which leads to the expression
\begin{equation}
\label{expression}
\R_{12}(\omega_L) = R_0 L \int_0^\infty dt \; n_2(L_2, t) e^{-t/\tau_s} \cos \omega_L t.
\end{equation}
Substituting Eq. (\ref{convolution}) into Eq. (\ref{expression}) and performing integration over
time with the help of the $\delta$-function, we get
\begin{widetext}
\begin{equation}
\label{double}
\R_{12}(\omega_L) =
 \frac{ R_0 (L_1 + L_2) l}{\tau_t \tau_s} \int_0^\infty dt_1 \int_0^\infty dt_2 \; P_{L_1}(t_1) P_{L_2}(t_2)
e^{-(t_1 + t_2)/\tau_s}\left( \cos\omega_L t_1 \cos\omega_L t_2 -
\sin\omega_L t_1 \sin\omega_L t_2 \right),
\end{equation}
We now notice that, for both terms in the brackets, the double integral Eq. (\ref{double})
factorizes into a product of single integrals, which, in turn, can be expressed through the
functions $R(\omega_L)$ and ${\tilde R}(\omega_L)$ for a {\em single} wire.     The final expression
for $\R_{12}(\omega_L)$ reads
\begin{equation}
\label{R12H}
\R_{12}=
\frac{\tau_s}{\tau_t} \frac{(L_1 +L_2) l}{L_1 L_2} \frac{1}{R_0}
\left(R(\omega_L, L_1) R(\omega_L, L_2) - {\tilde R}(\omega_L, L_1)
{\tilde R}(\omega_L, L_2) \right).
\end{equation}
Similarly, for orthogonal magnetizations of injector and detector we get
\begin{equation}
{\Rs}_{12}=
\frac{\tau_s}{\tau_t} \frac{(L_1 +L_2) l}{L_1 L_2} \frac{1}{R_0}
\left(R(\omega_L, L_1) \tilde{R}(\omega_L, L_2) + {\tilde R}(\omega_L, L_1)
R(\omega_L, L_2) \right).
\end{equation}
According to Eq. (\ref{R12H}), $\R_{12}$ depends on {\em both} the position of the bridge and the position
of the detector. In Appendix~A we demonstrate that the dependence of the
concentration, $n_2(L_2, t)$, on the position of the bridge drops out, so
that $\R_{12}$ only depends only on the distance, $L=L_1+L_2$ between the injector
and the detector.  This observation allows $L_2$ to be set to zero in Eq. (\ref{R12H});
correspondingly $L_1$ should be set equal to the total length $L$.
Technically, this implies that we can use the short-distance asymptotes, Eq. (\ref{smallL})
and Eq. (\ref{smallLtilde}), while for $R(\omega_L, L)$ and $\tilde{R}(\omega_L, L)$
the general expressions should be used.  These general expressions\cite{vanWeesPioneering,AnalyticalHanle} follow
from Eq. (\ref{f(y)}):
\begin{equation}
\label{propto}
R(\omega_L, L)
=\Bigl(\frac{\pi}{\vert y \vert}\Bigr)^{1/2}\exp{\Bigl[-2\vert y\vert^{1/2}\cos\frac{\phi}{2}\Bigr]}
\cos\Bigl(\frac{\phi}{2}+2\vert y\vert^{1/2}\sin\frac{\phi}{2}\Bigr),
\end{equation}
where the magnitude, $|y|$, and phase, $\phi$, are defined by Eq. (\ref{complex}).
The corresponding expression for $\tilde{R}(\omega_L, L)$ differs from Eq. (\ref{propto})
by the replacement of $\cos$ with $\sin$ in the second factor.

Summarizing, in the geometry of two wires with a bridge,  the Hanle curve measured by the
first detector is described by $R(\omega_L,L)$, Eq. (\ref{propto}), while the Hanle curve
measured by the second detector has the shape given by
\begin{equation}
\label{VeryLong}
\R_{12} \propto
\frac{\tau_s}{\tau_t} \frac{l}{L}
\left( \frac{\sqrt{\sqrt{1+\omega_L^2 \tau_s^2}+1}}{\sqrt{1+ \omega_L^2\tau_s^2}} R(\omega_L, L)  -\frac{\sqrt{\sqrt{1+\omega_L^2 \tau_s^2}-1}}{\sqrt{1+ \omega_L^2\tau_s^2}} {\tilde R}(\omega_L, L)
 \right).
\end{equation}
\end{widetext}
The most dramatic difference in the shapes of two Hanle curves emerges in the limit of
short wires, ${\cal L}\ll 1$. Substituting Eqs. (\ref{smallL}) and (\ref{smallLtilde}) into
Eq. (\ref{VeryLong}) we get the amusingly simple expressions for ${\cal R}_{12}$, $\tilde{\cal R}_{12}$
\begin{equation}
\label{surprising}
{\cal R}_{12}=\frac{1}{1+\omega_L^2\tau_s^2},~~~~\tilde{\cal R}_{12}=\frac{\omega_L\tau_s}{1+\omega_L^2\tau_s^2}.
\end{equation}
This means that, while the first detector measures the shape Eq. (\ref{smallL}),
the second detector measures a simple Lorentzian Eq. (\ref{optical}), as in optical measurements. In Fig.~\ref{figH}(a), this difference in shapes is illustrated graphically.
We see that ${\cal R}_{12}$ in the second wire is not only narrower, but also  possesses a distinctively different shape.
Fig. \ref{figH}(b) illustrates that the difference in the shapes of the two Hanle curves gradually vanished as the wires get longer. Qualitatively this can be understood from
Eq. (\ref{VeryLong}). In a long wire the Hanle curve is narrow. This allows us to set
$\omega_L\tau_s \ll 1$ in the prefactors in the brackets. Then the first prefactor
close to $1$, while the second prefactor is much smaller. Thus we conclude that
the ratio of $\R_{12}$ to $\R_{11}$ is approximately constant.

\section{Coupling of a wire to the loop}
\begin{figure}
\includegraphics[width=77mm]{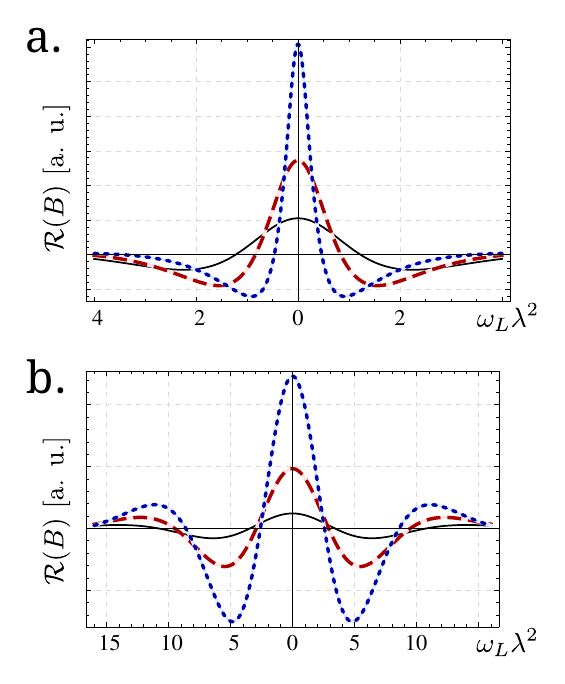}
\caption{[Color online] (a) Evolution of the Hanle curves, ${\cal R}(\omega_L)$,  with radius, $a$. The curves are
plotted from Eq. (\ref{long}) for three values of the dimensionless loop radius
$\lambda^{-1}=a/(D\tau_s)^{1/2}$. Dotted, dashed and solid lines correspond to $\lambda=1.7$,
$1$, and  $0.7$, respectively. (b) same as (a) for high-temperature domain $\lambda = 0.35$
(dotted), $0.32$ (dashed), and $0.27$ (solid). }
\label{figCircle}
\end{figure}
As a last example of the modification of the Hanle profile with restricted
geometry consider a loop tunnel-coupled to a wire, see Fig. \ref{devices}.
The injector is located in the wire while the detector is located in the loop.
The spin-transport equation has a form
\begin{equation}
\label{azimuthal}
\frac{\partial }{\partial t} {\bm S}(\theta, t)
 =  {\bm \omega}_L \times {\bm S}(\theta, t) +  \frac{D}{a^2} \frac{\partial^2}{\partial \theta^2} {\bm S}(\theta,t)
 - \frac{{\bm S}(\theta, t)}{\tau_s},
\end{equation}
where $\theta$ is the azimuthal coordinate and  $a$ is the radius of the loop.
This equation also allows a factorization: ${\bm S}(\theta,t)={\bm s}(t)n(\theta,t)$,
where ${\bm s}(t)$ satisfies  Eq. (\ref{dynamics}), while the equation for $n(\theta,t)$
reads
\begin{equation}
\label{n-circle}
\frac{\partial n}{\partial t} = \frac{D}{a^2} \frac{\partial^2 n}{\partial \theta^2}.
\end{equation}
The solution of this equation satisfying the initial condition, $n(\theta, 0) = \delta(\theta)$, can be presented as a sum of angular harmonics
\begin{equation}
\label{harmonics}
n(\theta, t) = \frac{1}{2 \pi} + \frac{1}{\pi} \sum_{k=1}^\infty
\exp\left[ -\frac{Dk^2t}{a^2} \right] \cos(k \theta).
\end{equation}
Assuming that the detector is located at $\theta=\pi$, we find the following
expression for the spin density
\begin{equation}
\label{sum0}
S_x(t) = \frac{1}{2\pi} \sum_{k=-\infty}^{\infty}
(-1)^k\exp\left[-\frac{Dk^2t}{a^2} -\frac{t}{\tau_s}\right]
\cos\omega_L t.
\end{equation}
Integration of $S_x(t)$ over time yields the nonlocal resistance
in the form of the infinite sum
\begin{equation}
\label{sum}
R(\omega_L) = R_0  \sum_{k=-\infty}^{\infty}
\frac{ 1 + \dfrac{D k^2 \tau_s}{a^2}}
{ \left( 1 + \dfrac{D k^2 \tau_s}{a^2} \right)^2 + \omega_L^2 \tau_s^2}.
\end{equation}
Closed expression for $R(\omega_L)$ can be obtained with the use of the identity
\begin{multline}
\label{long}
\sum_{k=-\infty}^{\infty}\frac{\left(-1\right)^k(1+k^2\lambda^2)}{(k^2\lambda^2+1)^2+z^2}
=\frac{2}{\left(1+z^2\right)^{1/2}} \\
\times\Biggl[\frac{x\sinh(x)\cos(y) -
y\sin(y) \cosh(x)}
{\cosh 2x-\cos 2y}\Biggr],
\end{multline}
%
%
where
\begin{equation}
\label{x,y}
x=\frac{\pi}{\lambda} \sqrt{\dfrac{\sqrt{1+z^2}+1}{2}},~~y= \frac{\pi}{\lambda} \sqrt{\dfrac{\sqrt{1+z^2}-1}{2}}.
\end{equation}
Expressing of nonlocal resistance with the help of Eq. (\ref{long}) requires the following identifications
\begin{equation}
z=\omega_L\tau_s,~~~~\lambda=\frac{(D\tau_s)^{1/2}}{a}.
\end{equation}
The  prefactor $\pi/\lambda$ in Eq. (\ref{x,y}) is equal to $2\pi a/\left(4D\tau_s\right)^{1/2}$, which is
the circumference of the loop in the units of
spin-diffusion length. The identity Eq. (\ref{long}) suggests that nonlocal resistance
depends on parameter $y$ in an oscillatory fashion. To clarify the physical meaning of these oscillations consider the limit of weak fields, $z\ll 1$, so that the precession angle
 of spin, $\delta \varphi$, during the time $\tau_s$ is small.
 In this limit the parameter $y$ can be cast in the form
 \begin{equation}
 \label{approximate}
 y\,\big|_{z\ll 1}\approx \frac{2\pi a}{\left(D\tau_s\right)^{1/2}}\delta\varphi.
 \end{equation}
The first factor in Eq. (\ref{approximate}) can be interpreted as a number of intervals, each having the length equal to the spin--diffusion length, covered by electron before it makes a full loop. Since the
spin is rotated by $\delta\varphi$ over  each interval, the parameter $y$ can be interpreted as
a full rotation angle for the whole loop.
Then the periodicity of nonlocal resistance corresponds to this full angle
being $\pi$, $2\pi$, and so on.

To interpret the oscillations in strong fields, $z\gg 1$, we rewrite the parameter $y$ as
\begin{equation}
 \label{approximate1}
 y\,\big|_{z\gg 1}\approx \frac{2\pi a}{\left(D/\omega_L\right)^{1/2}}.
 \end{equation}
The denominator in Eq. (\ref{approximate1}) has the meaning of the length traveled during
one Larmour precession period. The fact that ${\cal R}$ is sensitive to  whether the circumference contains integer or half-integer number of these lengths can be interpreted as
an effect of finite step-size in the random walk.  Naturally, these oscillations are suppressed exponentially, since at strong fields we have $x\approx y$.

Suppose now that the temperature is low, so that $\tau_s$ is long. This means that, before the spin orientation is forgotten, the particle performs many loops, so that density $n(\theta)$ is nearly homogeneous. This, in turn, suggests that the Hanle profile is unaffected by the diffusion, and has a Lorentzian shape. Plotting ${\cal R}(\omega_L)$ from Eq. (\ref{long})
indicates that Lorentzian shape is achieved only for very small loops, such that $\lambda \gtrsim 40$. For moderate values of $\lambda \sim 1$, i.e.  for higher temperatures, the Hanle profiles are non-Lorentzian,  but rather resemble ${\cal R}(\omega_L)$ for a long wire, as illustrated in Fig. \ref{figCircle}(a). Finally for ``high" temperatures corresponding to
$\lambda \sim 0.3$, see Fig. \ref{figCircle}(b), the Hanle curves develop oscillations discussed above, while the magnitude of ${\cal R}(\omega_L)$ drops rapidly with $\lambda$.

\section{Discussion}
\begin{itemize}

\item

The fact that experimental Hanle curves are amazingly robust
motivated us to investigate whether the charge-transport characteristics could be inferred from their shapes. Namely,
in the system of two coupled wires, the tunneling time, $\tau_t$,
between the wires is a parameter which does not depend on spin.
Still, as it is seen in Fig. \ref{figTunnel},  the Hanle
curves calculated for a given wire, ${\cal R}_{11}$, and between
the wires, ${\cal R}_{12}$, have visibly different widths. The difference in widths is governed by the ratio. $\Delta_t/\Delta_s$, of fundamental ``band-structure"  parameters. Thus, this ratio can be
inferred from the comparison of these widths. Also, $\Delta_s$ can, in principle, be inferred independently from the shape of ${\cal R}_{11}$. Possibility of such an extraction of tunnel splitting is
facilitated by the fact that $\tau_s$ falls off with increasing temperature dramatically (as $T^{-3}$ for the Dyakonov-Perel mechanism \cite{dyakonov-perel}), whereas $\tau_t$ varies slowly.
This rapid change of $\tau_s$ with temperature allows for
a ``dimensional crossover" between 0D and 1D statistics of
diffusion paths within the same ring-shaped sample, see Fig. \ref{devices}(c). This crossover manifests itself not only in the width but also in the shape of  Hanle curve, which becomes a Lorentzian
at low temperatures.

\item

In a sense, our quest to reveal the statistics of diffusion
paths through the Hanle curves
is in line with  attempts taken  to
unravel this statistics from the
weak-localization correction to the
conductivity, $\Delta\sigma$ of a 2D sample\cite{Minkov,Minkov1}.

The Hanle profile comes from multiplying the diffusive
propagator by $\cos\omega_L t$ and integrating over
time. Similarly, the expression for $\Delta\sigma$ comes
from multiplying the diffusive return probability by
$\cos2\pi\Phi(t)$ where $\Phi(t)$ is the flux (in the unit of
the flux quantum) through the area covered by the diffusing particle after time $t$ and integrating over $t$.
Thus, both ${\cal R}$ and $\Delta\sigma$ are essentially the Fourier
transforms of the diffusion propagator. The role of the spin-flip time $\tau_s$ in spin transport is played by the phase-breaking time in
magnetoresistance.

In fact, a rapid decay of the phase-breaking time with temperature
was also exploited previously in the transport studies\cite{crossover0,crossover}  to demonstrate the
dimensional crossover from quasi-2D to purely 3D diffusion.
It should be noted, however, that while weak-localization relying
on the spatial coherence of electron shows up only at low temperatures, the Larmour precession survives at high temperatures
and gives rise to the Hanle curve.

\item
Throughout the paper we considered a two-wire geometry.
Another class of structures to which our results might be  applicable
is tunnel-coupled graphene layers.
It was previously demonstrated\cite{graphene2007,graphene2013,graphene2014,graphene2009,WeesGraphene} that a single  layer of
graphene can be used as a channel for nonlocal spin-transport measurements.
A possibility to fabricate two tunnel-coupled layers was also demonstrated
very recently.\cite{GeimScience2012,GeimTwo-Layer2012,two-layer2013,KoreanVertical}
The structures\cite{GeimScience2012,GeimTwo-Layer2012,two-layer2013,KoreanVertical} were fabricated in order
to realize the vertical gate-controlled graphene heterostructures.

\item
We considered the spin-current distribution
in a loop geometry.
Very recently\cite{Spin-Transport-Loop} a
measurement of nonlocal spin transport
in a loop geometry has been reported.
The importance of findings of
Ref. \onlinecite{Spin-Transport-Loop}.
is that the result of conversion of a spin current
into a charge current was revealed not through the voltage
buildup in an open-circuit geometry but rather by directly
measuring the circulation of current in the loop.

\item
The bridge between two channels shown in Fig. \ref{devices}(b) can be viewed
as a boundary condition for the diffusion equation that changes the random-walk trajectories leading to a modified shape of the Hanle curves. The origin of
such a boundary condition can be simply a finite length of the channel. This
situation was recently considered theoretically\cite{1D-0D}. In accord with our findings, the result of decreasing the channel length is the crossover of the Hanle shape to a Lorentzian.

\end{itemize}

\section{Acknowledgements}

 This work was supported by NSF through MRSEC DMR-1121252. E.M. acknowledges
 support from the Department of Energy, Office of Basic Energy Sciences,
 Grant No. DE-FG02-06ER46313.

\begin{widetext}
\begin{appendix}
\section{~}
In order to substantiate the statement made in Sect. III that the concentration
profile in the second wire,
\begin{equation}
\label{n2}
n_2(L_2, t) \propto \int_0^t dt_1 \; P_{L_2}(t - t_1) P_{L_1}(t_1),
\end{equation}
does not depend on the position, $L_1$, of the bridge, it is convenient
to use the Fourier representation of the diffusive propagators $P_{L_1}$ and $P_{L_2}$.
In this representation Eq. (\ref{n2}) acquires the form
\begin{equation}
n_2  \propto \int_0^t dt_1 \; \left(  \int\frac{dq_2}{2 \pi}
\exp\left[ -Dq_2^2(t-t_1) + i q_2 L_2 \right]
\int\frac{dq_1}{2\pi} \exp\left[ -Dq_1^2t_1 + i q_1 L_1 \right] \right).
\end{equation}
Performing the time integration, we get
\begin{equation}
\label{a3}
n_2  \propto \int\frac{dq_1}{2 \pi}  \int\frac{dq_2}{2 \pi}
\exp\left[i q_1 L_1 + i q_2 L_2  \right] \left( \frac{\exp[-Dq_1^2 t] - \exp[-Dq_2^2 t]}{q_1^2 - q_2^2} \right).
\end{equation}
The sum, $L=L_1+L_2$, which is the total length, does not depend on the position of the bridge, while the difference $l=L_1-L_2$ is fully determined by the position of the
bridge.  In order to decouple $L$ and $l$, we introduce new variables
\begin{equation}
u = q_1 + q_2, \quad v = q_1 - q_2,
\end{equation}
so that the integral Eq. (\ref{a3}) assumes the form
\begin{equation}
\label{a5}
n_2  \propto \int \frac{du}{u} \exp\left[ -\frac{Dtu^2}{4} + i \frac{L}{2}u \right]
\int \frac{dv}{v} \exp\left[ -\frac{Dtv^2}{4} + i \frac{l}{2}v \right]
\sinh\left(\frac{Dtuv}{2}\right).
\end{equation}
Independence of $n_2$ on the position of the bridge implies that
$ \partial n_2/\partial l =0$.  Differentiating Eq. (\ref{a5}) with
respect to $l$, we get
\begin{equation}
\label{a6}
\frac{\partial n_2}{\partial l}  \propto \int \frac{du}{u} \exp\left[ -\frac{Dtu^2}{4} + i \frac{L}{2}u \right]
\int dv \; \exp\left[ -\frac{Dtv^2}{4} + i \frac{l}{2}v \right]
\sinh\left(\frac{Dtuv}{2}\right).
\end{equation}
Note that the internal integral in Eq. (\ref{a6}) can be readily evaluated
\begin{equation}
\int dv \; \exp\left[ -\frac{Dtv^2}{4} + i \frac{l}{2}v \right]
\sinh\left( \frac{Dtuv}{2} \right)
\propto \exp\left[ \frac{Dt u^2}{4}\right] \sin\frac{lu}{2}.
\end{equation}
This allows us to express $\partial n_2/\partial l$ as a single integral
\begin{equation}
\frac{\partial n_2}{\partial l} \propto \int \frac{du}{u} \exp\left[ \frac{iLu}{2} \right]
\sin\frac{l u}{2} =
\frac{1}{2} \int du \frac{ \sin\frac{(L+l)u}{2} -\sin\frac{(L-l)u}{2}}{u}.
\end{equation}
From the identity
\begin{equation}
\int\limits_{-\infty}^{\infty} \frac{ds}{s} \sin \alpha s = \pi\, \text{sign}(\alpha),
\end{equation}
we conclude that indeed $\partial n_2/\partial l$ is zero {\em as long as $l < L$}.
Thus the concentration, $n_2(L_2, t)$, does not depend on the position of the bridge
only when the bridge is located between the injector and the detector.
\end{appendix}
\end{widetext}

\end{document}